# A HUMAN VISUAL SYSTEM-BASED 3D VIDEO QUALITY METRIC

*Amin Banitalebi-Dehkordi[1], Student Member, IEEE, Mahsa T. Pourazad[1,2], Member, IEEE, and Panos Nasiopoulos[1], Senior Member, IEEE*

[1]Department of Electrical & Computer Engineering, University of British Columbia, Canada
[2]TELUS Communications Inc., Canada
{dehkordi, pourazad, panosn}@ece.ubc.ca

## ABSTRACT

Although several 2D quality metrics have been proposed for images and videos, in the case of 3D efforts are only at the initial stages. In this paper, we propose a new full-reference quality metric for 3D content. Our method is modeled around the HVS, fusing the information of both left and right channels, considering color components, the cyclopean views of the two videos and disparity. Performance evaluations showed that our 3D quality metric successfully monitors the degradation of quality caused by several representative types of distortion and it has 86% correlation with the results of subjective evaluations.

## 1. INTRODUCTION

Assessing the quality of 3D content is much more difficult than that of 2D content. Existing 2D quality metrics cannot be directly used to evaluate 3D quality since they do not take into account the effect of depth and the binocular properties of the human visual system (HVS). Such efforts result in low correlation between the objective and the subjective test results. An example of this is the proposed approach in [1], where existing 2D quality metrics are applied to the right and left views separately, and the results are averaged over two views to quantitatively evaluate the quality of the 3D picture.

To account for the effect of depth when evaluating the quality of 3D content, researchers in [2] proposed a 3D quality metric that is based on the temporal and spatial variance of the disparity and the motion in the scene. While the proposed metric takes into account the effect of depth, it does not include the effect of 2D associated quality factors such as contrast and sharpness. In addition, this method uses MSE (Mean Square Error) as a quality measure, which does not accurately represent the human visual system. Another group of researchers categorized the 3D content distortions to monoscopic and stereoscopic types and proposed separate metrics for each type of distortions [3]. In this approach, while the monoscopic quality metric quantitatively measures the distortions caused by blur, noise and contrast-change, the stereoscopic metric exclusively measures the distortions caused by depth inaccuracy. The main drawback of this approach is that it is unable to accurately measure the overall 3D quality as it does not attempt to fuse the 2D and 3D associated factors. In [4] the authors propose a 3D quality metric, which is a combination of 2D quality metrics with disparity map quality. However this method fails to include the color components. Another shortcoming is that the cyclopean view quality is not taken to account. Recently, a full-reference quality metric was proposed in [5], which is designed specifically for mobile 3DTV content. Although this metric addresses the shortcomings of the 2D quality metrics, it is still strongly dependent on MSE for block matching and measuring the similarity between block structures and between depth maps. Moreover, its performance is only verified for compressed videos.

In this paper, we propose a new full-reference 3D quality metric, which includes the quality of the fused right and left views (cyclopean view) similar to human visual system and combines it with the quality of each of the views, and the quality of depth maps. The proposed metric can be tailored for different applications as it takes into account the display size, video resolution and the distance of the viewer from the display. The performance of the proposed metric is validated by subjective tests, using 4 reference and 40 modified videos and 19 subjects, following the ITU-R BT.500-11 recommendation.

The rest of this paper is organized as follows: Section 2 is dedicated to the description of the proposed metric. Subjective tests are presented in section 3 while the results and discussions are provided in section 4. Section 5 concludes the paper.

## 2. PROPOSED 3D QUALITY METRIC

Our proposed Human-Visual-system-based 3D (HV3D) quality metric takes into account the quality of individual views, the quality of the cyclopean view, as well as the quality of the depth information as follows:

$$HV3D = w_1 Q_{R'} + w_1 Q_{L'} + w_2 Q_{R'L'} + w_3 Q_{D'} \quad (1)$$

where $Q_{R'L'}$ is the quality of the cyclopean view (fusion of the right and left view), $Q_{R'}$ and $Q_{L'}$ are the quality of the distorted right and left views respectively compared to the reference views, $Q_{D'}$ is the quality of the depth information of distorted views, and $w_1$, $w_2$, and $w_3$ are weighting constants. The weighting constants are chosen so that the quality measures used in our method are given different importance in order to lead to the best possible results. Fig. 1 illustrates the flow chart of the proposed method. The following subsections elaborate on the different quality measures used in our method. The weighting constants are decided based on subjective tests presented in Section 3.

### 2.1. Quality of Individual Views

Our proposed HV3D quality metric considers the independent quality of each view compared to its corresponding reference view as shown in equation (1). The quality of the distorted right view with respect to its matching reference view is calculated as follows:

$$w_1 Q_{R'} = w_1 VIF(Y_R, Y_{R'}) + w_4 VIF(U_R, U_{R'}) + w_4 VIF(V_R, V_{R'}) \quad (2)$$

where $Y_R$ and $Y_{R'}$ are luma information of the reference and distorted right views respectively, $U_R$ and $V_R$ are the chroma information of the reference right-view, $U_{R'}$ and $V_{R'}$ are the chroma information of the distorted right-view, $w_1$ and $w_4$ are weighting constants, and VIF is the visual information fidelity index (refer to [6] for more details). Note that since the sensitivity of the human visual system is different for luminance and chrominance, different weighting constants are assigned to them. The quality of the left view is calculated in the same fashion.

### 2.2. Quality of Cyclopean View

One of the most amazing properties of the human stereo vision is the fusion of the left and right views of a scene into a single cyclopean view. To comply with the human visual system, we include the quality of the cyclopean view in our proposed quality metric as well (please see equation 1). Note that the quality of the cyclopean view is different from the quality of the two individual views, To measure the quality of the cyclopean view, first we imitate the binocular vision and form the cyclopean view by combining the corresponding areas from the left and right views. This is done by finding the matching blocks between right and left views (applying a block matching technique). To this end, the luma information of each view is divided into 16×16 blocks. Note that the block size is chosen for HD video applications, to significantly reduce the complexity of our approach (i.e., search, matching and variance) while allowing efficiently extracting the local structural similarities between views. For SD video applications 8x8 blocks may be used.

In order to model the cyclopean view, once the matching blocks are detected, the information of matching blocks in the left and right views needs to be fused. Here we apply the 3D-DCT transform to each pair of matching blocks (left and right views) to generate two 16×16 DCT-blocks which contain the DCT coefficients of the fused blocks. Since the human visual system is more sensitive to the low frequencies of the cyclopean view [5], we only keep the first level of coefficients which is a 16×16 DCT-block and discard the other ones.

Another property of the human visual system is its sensitivity to contrast. To take into account this feature, we derive a 16×16 Contrast Sensitivity Function (CSF) modeling mask and apply it to the 16×16 DCT-block so that the frequencies that are of more importance to the human visual system are assigned bigger weights. This is illustrated as follows:

$$XC = \sum_{i=1}^{16} \sum_{j=1}^{16} C_{i,j} X_{i,j} \quad (3)$$

where $XC$ is our cyclopean-view model for a pair of matching blocks in the right and left views, $X_{i,j}$ are the low-frequency 3D-DCT coefficients of the fused view, $i$ and $j$ are the horizontal and vertical indices of coefficients, and $C_{i,j}$ is our CSF modeling mask. $C_{i,j}$ is derived based on the idea presented in [7] which utilizes JPEG quantization table and creates a mask, which includes larger coefficients for more visually important DCT elements. To derive a mask similar to [7], we adopt the 8x8 JPEG quantization table and create an 8x8 mask. Note that the JPEG quantization tables have been obtained from a series of psychovisual experiments to determine the visibility thresholds for the DCT basis functions. Based on this, the DCT coefficients that the human visual system is more sensitive to, are quantized less than the other coefficients so that the more visually important content is preserved during the course of compression. In our application, instead of quantizing coefficients we would like to scale the DCT values so that bigger weights are assigned to visually important content (see equation (3)). To this end, our CSF mask is designed such that the ratio among its coefficients is inversely proportional to the ratio of the corresponding elements in the quantization table of JPEG. Since this mask needs to be applied to 16×16 3D-DCT blocks, cubic interpolation is used to up-sample the coefficients of the mask and create a 16×16 mask. Moreover, the elements of this mask are selected such that their average is equal to one. This guarantees that, in the case of uniform distortion distribution, the quality of each block within the distorted cyclopean view coincides with the average quality of the same view. Once we obtain the cyclopean-view model for all the blocks within the distorted and reference 3D views, the quality of the cyclopean view is calculated as follows:

$$Q_{R'L'} = VIF(D,D')^\beta \sum_{i=1}^{N} \frac{SSIM(IDCT(XC_i),IDCT(XC'_i))}{N} \quad (4)$$

where $XC_i$ is the cyclopean-view model for the $i^{th}$ matching block pair in the reference 3D view, $XC'_i$ is the cyclopean-view model for the $i^{th}$ matching block pair in the distorted 3D view, $IDCT$ stands for inverse 2D discrete cosine transform, $D$ is the depth map of the reference 3D view, $D'$ is the depth map of the distorted 3D view, $N$ is the total number of blocks in the each view, $\beta$ is a constant, and SSIM is the structural similarity metric designed specifically for measuring similarities between two images [8]. SSIM compares the structure of two images only and does not take into account the effect of small geometric distortions. In equation (4), small geometric distortions between right and left images (within the cyclopean view) are reflected in the depth map, thus the VIF index of the depth map of the distorted video with respect to the original video's depth map is used as a scaling factor in conjunction with SSIM. This allows measuring the quality of the cyclopean view more accurately. Considering that geometrical distortions are the source of vertical parallax, which causes severe discomfort for viewers, $\beta$ in the equation (4) is empirically assigned to 0.7 (resulted from a series of subjective tests), so that more importance is given to VIF index (the range of VIF and SSIM metrics by default is between 0 to 1).

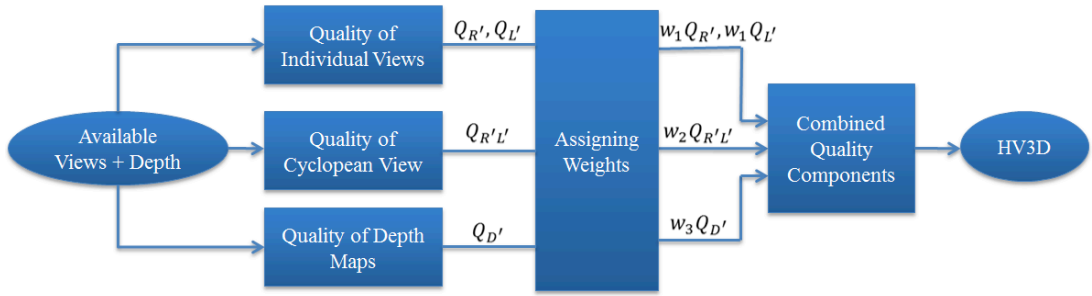

**Fig. 1.** Flowchart of HV3D

## 2.3. Quality of Depth Maps

The quality of depth information plays an important role in the perceptual quality of the 3D content, and for this reason it has been included in our 3D quality metric as denoted in equation (1). The quality of the depth map becomes more important if there are several different depth levels in the scene. On the contrary, in a scene with limited number of depth levels, the quality of the depth map plays a less important role in the overall 3D quality. Given the above observations, in our approach we chose the quality of the depth map to be variance-dependent and we formulated it as follows:

$$Q_{D'} = VIF(D,D')^\beta \sum_{i=1}^{N} \frac{\sigma_{d_i}^2}{N.\max(\sigma_{d_i}^2 \mid i=1,2,..,N)} \quad (5)$$

where $\sigma_{d_i}$ is the variance of block $i$ in the depth map of the 3D reference view, $\beta$ is a constant = 0.7 and $N$ is the total number of blocks. The local disparity variance, $\sigma_{d_i}^2$, is defined as follows:

$$\sigma_{d_i}^2 = \frac{1}{64 \times 64 - 1} \sum_{k,l=1}^{64} (M_d - R_{k,l})^2 \quad (6)$$

where $M_d$ is the mean of the depth values of each 64x64 block (outer block around the $i^{th}$ 16×16 block) in the normalized reference depth map. Here local disparity variance is calculated over a block size area of 64x64, since this is the area that can be fully projected onto the eye fovea when watching a 46" HD 3D display from a typical viewing distance of 3 meters. The reference depth map has been normalized with respect to its maximum value in each frame, so that the depth values range from 0 to 1. $R_{k,l}$ is the depth value of pixel ($k, l$) in the outer 64x64 block within the normalized reference depth map.

## 3. SUBJECTIVE TESTS AND WEIGHTING CONSTANTS

To find the weighting factors for our proposed HV3D quality metric and validate its performance, we performed subjective tests using two different 3D databases (one set for training and one set for validation). These sequences are selected from the test videos in [1] and the 3D video database of the Digital Multimedia Lab (DML) at the University of British Columbia, such that the datasets contain videos with fast motion, slow motion, dark and bright scenes, human and non-human subjects, and a wide range of depth. The specifications of the test videos are summarized in TABLE I. Four different types of distortions are applied to each of these eight stereo-videos. The following are distortions commonly used to evaluate the performance of quality metrics. The level of distortion in each case is such that it leads to visible artefacts that in turn allow us to correlate subjective tests - Mean Opinion Score (MOS) - with objective results.

1) White Gaussian noise with the zero mean and variance of 0.01 for all videos.
2) Compression distortions. The test videos are encoded using the emerging HEVC standard (HM software version 5.0). The low delay configuration setting with the GOP size of 4 was used. In this case, two levels of distortion were used. The quantization parameter (QP) was set at 35 and 40 to investigate the performance of our proposed metric at two different compression-distortion levels with visible artifacts.
3) Gaussian low pass filter with the size of 4 and standard deviation of 4.
4) Shifted (increased) the mean intensity of videos by 20 (out of 255).

TABLE I
DATASETS

| | Sequence | Resolution | Frame Rate (fps) | Number of Frames |
|---|---|---|---|---|
| Training set | Soccer2 | 1080×1920 (upsampled from 480x720) | 30 | 450 |
| | Flower | 1080×1920 (upsampled from 270x480) | 30 | 112 |
| | Horse | 1080×1920 (upsampled from 270x480) | 30 | 140 |
| | Car | 1080×1920 (upsampled from 270x480) | 30 | 235 |
| Validating set | Cokeground | 1080×1920 | 30 | 210 |
| | Ball | 1080×1920 | 30 | 150 |
| | Alt-Moabit | 1080×1920 (upsampled from 384x512) | 30 | 100 |
| | Hands | 1080×1920 (upsampled from 270x480) | 30 | 251 |

After applying the five above-mentioned distortions to the eight stereo videos, we obtain 40 distorted stereo videos. Subjective evaluations were performed using the training set (see TABLE I) to tune the weighting parameters in our HV3D metric as denoted in the equation (1). Subsequent subjective tests were performed to evaluate our 3D quality metric using the validating dataset. The viewing conditions for subjective tests were set according to the ITU-R Recommendation BT.500-11 [9]. Twenty-one observers participated in our subjective tests, ranging from 21 to 29 years old. All subjects had none to marginal 3D image and video viewing experience. They were all screened for color and visual acuity (using Ishihara and Snellen charts), and also for stereo vision (Randot test – graded circle test 100 seconds of arc). The evaluation was performed using a 46" Full HD Hyundai 3D TV (Model: S465D) with passive glasses. The TV settings were as follows: brightness: 80, contrast: 80, color: 50, R: 70, G: 45, B: 30. The 3D display and the settings were based on the MPEG recommendations for the subjective evaluation of the proposals submitted in response to the 3D Video Coding Call for Proposals [10].

After a short training session, the viewers were shown the distorted and reference stereoscopic test sequences (each 10 seconds long) in random order, so that the viewer would watch the reference and the distorted versions of the same sequence consecutively without knowing which video is the reference one. Between test videos, a four-second gray interval was provided for allowing the viewers to rate the perceptual quality of the content and relax their eyes before watching the next video. Here, the perceptual quality reflects whether the displayed scene looks pleasant in general.

In particular, subjects were asked to rate a combination of "naturalness", "depth impression" and "comfort" as suggested by [11]. There were 10 quality levels (1-10) for ranking the videos, where score 10 indicated the highest quality and 1 indicated the lowest quality. After collecting the experimental results, we removed the outliers based on the ITU-R Recommendation BT.500-11 (there were two outliers) and then the mean opinion scores (MOS) from viewers were calculated.

The depth maps of the reference and the distorted stereo videos were estimated using the depth estimation reference software provided by MPEG for 3D Video activities [12]. To find the weighting factors, similar to the studies in [5] and [13], the least mean square technique was used such that the difference between the HV3D metric values and the MOS values was minimized for the training video set. Our objective is to determine the best values for the weighting constants $w_1$, $w_2$, $w_3$ and $w_4$ which result in the minimum mean of square errors between our HV3D index and the MOS. TABLE II shows the resulting values for these constants.

Our HV3D quality metric is taking values between 0 and 1 (because it is optimized to be correlated with MOS/10), and higher than 1 in case quality is improved.

TABLE II
WEIGHTING CONSTANTS

| $w_1$ | $w_2$ | $w_3$ | $w_4$ |
|---|---|---|---|
| 0.14 | 0.1208 | 0.05 | 0.1353 |

## 4. RESULTS AND DISCUSSION

Performance evaluation of our HV3D metric was tested (based on the resulted coefficients) using the validation dataset. Fig. 2 shows the relationship between the MOS and the resulting values from our quality metric for the entire validation set and all 5 different distortions (as described in section 3). A logistic fitting curve is used to more clearly indicate the correlation between subjective results (MOS) and the results derived by the objective metric. As it can be observed, our HV3D objective metric manages to clearly distinguish the resulting visual quality degradation. The Spearman ratio between the MOS and the HV3D objective quality metric over the whole validation dataset for all four representative kinds of distortion is 0.8601, which shows strong statistical dependency between the subjective and objective results.

In our future work, the performance of our 3D objective quality metric is compared with that of other existing metrics. Moreover we investigate the performance of our metric for mobile 3D TV applications as well as 3D video coding applications.

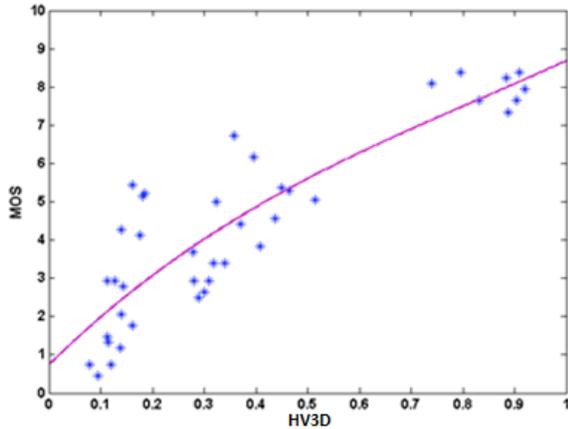

**Fig. 2.** Comparing the subjective results with objective results using HV3D objective quality metrics

## 5. CONCLUSION

In this paper we proposed a new quality metric for 3D video. Our approach combined the quality of each of the views, the quality of the cyclopean view and the quality of depth maps. Performance evaluations over the validation dataset and subjective test results revealed the fact that the proposed quality metric has almost 86% correlation with subjective test results.